\pdfoutput=1
\documentclass[11pt]{article}

\usepackage[margin=1.1in]{geometry}
\usepackage{amsmath,amssymb}
\usepackage{booktabs}
\usepackage{graphicx}
\usepackage{url}
\usepackage[hidelinks]{hyperref}
\usepackage{xcolor}
\usepackage[numbers,sort&compress]{natbib}
\usepackage{listings}
\usepackage{microtype}

\newcommand{\pkg}[1]{\textsf{#1}}
\newcommand{\code}[1]{\texttt{#1}}

\title{Wrap or Rewrite? Reviving the 1989 \emph{GlobalMinimum} Bayesian
Global-Optimization Library for R and Python}

\author{Audris Mockus\\
  University of Tennessee, Knoxville\\
  \texttt{audris@utk.edu}}

\date{\today}

\begin{document}
\maketitle

\begin{abstract}
The \emph{GlobalMinimum} (MINIMUM) library of Jonas Mockus (1989)
implements some of the earliest practical Bayesian global-optimization
methods---one-step Bayesian search over a simple surrogate, Wiener-process
one-dimensional models, low-discrepancy designs, and clustering
multistart---that anticipate the acquisition-function machinery of modern
Bayesian optimization. We modernize this Fortran~66/77 codebase along the
two standard routes: \emph{wrapping} the original compiled routines behind
idiomatic R and Python packages, and \emph{translating} the core methods to
Rust. The exercise yields three contributions. First, we release polished R
and Python packages in which every optimizer runs the genuine 1989
algorithms (with callback objectives, deterministic seeding, full
evaluation traces, and input validation that retrofits memory- and
I/O-safety onto COMMON-block Fortran), plus a Rust crate exposing both the
translation and the wrapped original. Second, we dissect a
translation-fidelity incident: the Rust port of the one-step Bayesian
planner was up to $12\times$ slower than the wrapped Fortran---an artifact
initially misattributed to foreign-function overhead. Controlled
decomposition shows call overhead is microseconds and near-identical across
backends; the real cause was two dropped pruning rules (a partial-distance
early exit and a second-minimum bound) whose restoration recovers bitwise
trajectory agreement \emph{and} speed parity. The same audit surfaced
transposed coefficient matrices in re-implemented Shekel and Hartmann test
functions---quiet correctness bugs of the rewrite route. Third, we
benchmark the 1989 methods against modern global optimizers in both
ecosystems (SciPy, NLopt, CMA-ES, Gaussian-process Bayesian
optimization, GenSA, DEoptim, GA) on a COCO-style
shifted-instance suite of 16~test families at dimensions 2--10 under
matched evaluation budgets, using performance and data profiles, with
external validation on the official BBOB/COCO suite; the wrapped
methods replicate bitwise across the two ecosystems. On the classical
test set the coordinate Wiener-model search EXKOR is competitive with
the best modern methods (solving 56\% of instances to gap $10^{-2}$,
statistically tied with SHGO, dual annealing, and GenSA); on BBOB,
whose functions are predominantly non-separable and ill-conditioned,
it drops to mid-field (15\% versus 36\% for dual annealing), cleanly
quantifying how much of that strength is separability. At tight
budgets modern annealing and Gaussian-process expected improvement
dominate the 1989 acquisition machinery decisively---the latter at
three orders of magnitude more solver-internal time per run. Finally,
we ask whether the historic components compose: EXKOR/A2, which routes
EXKOR through the eigenframe estimated by the library's own screening
routine, repairs four of EXKOR's five zero-solve families and improves
its mean rank in both ecosystems, at the cost of a statistical tie in
aggregate solve rate---composition buys robustness, not throughput.
All code, data, and scripts are available for replication.

\end{abstract}

\section{Introduction}
\label{sec:intro}

Between the first expected-improvement formulations of
Kushner~\cite{kushner1964new} and the modern Gaussian-process machinery
popularized by EGO~\cite{jones1998efficient} and surveyed
in~\cite{shahriari2016taking,frazier2018tutorial} lies a substantial body
of practical Bayesian global-optimization software developed by Jonas
Mockus and collaborators in Vilnius from the 1960s through the
1980s~\cite{mockus1978application,mockus1989bayesian}. Its concrete
artifact, the \emph{MINIMUM}/\emph{GlobalMinimum} Fortran
library~\cite{globalminimumfortran} distributed with the 1989
monograph~\cite{mockus1989bayesian}, packages fourteen optimization and
analysis routines: the one-step Bayesian method \code{BAYES1}, a
Wiener-model one-dimensional global search \code{EXTR} and its
coordinate-descent extension \code{EXKOR}, uniform-search and clustering
multistart methods (\code{UNT}, \code{GLOPT}), LP-$\tau$ low-discrepancy
search \code{LPMIN}~\cite{sobol1967distribution}, local methods
(\code{MIVAR4}, \code{FLEXI}, \code{REQP}), Monte-Carlo baselines
(\code{MIG1}, \code{MIG2}), and variable-screening procedures
(\code{ANAL1}, \code{ANAL2}).

This library is scientifically interesting today for two reasons. As
\emph{methods}, its algorithms are extremely cheap
surrogate-based optimizers: the \code{BAYES1} planner selects each new
point by maximizing a closed-form acquisition over the full evaluation
history at $O(\ell n)$ cost per candidate, in contrast to the $O(\ell^3)$
Gaussian-process algebra of modern Bayesian optimization. How such methods
fare against contemporary default choices (differential evolution,
generalized simulated annealing, DIRECT, CMA-ES) under matched budgets is
an empirical question that, to our knowledge, has not been answered with
modern benchmarking methodology. As \emph{software}, the library is a
representative specimen of scientific legacy code---fixed-form Fortran with
COMMON-block state, link-time objective binding, single-precision
utilities promoted by compiler flag, and console I/O woven into numerical
routines---and thus a concrete case study in the perennial engineering
question: \textbf{wrap the original, or rewrite it in a modern language?}

We did both, and instrumented the difference. This paper reports:

\begin{itemize}
\item \textbf{Packages} (Section~\ref{sec:packages}). A CRAN-style R
  package in which all fourteen routines run the \emph{original}
  compiled Fortran, and a PyPI-style Python package wrapping the twelve
  optimization routines (the two screening procedures remain
  translation-backed in Python). A small trampoline (the upstream
  objective \code{FI(X,N)} is a link-time symbol, not an argument) and a
  language-agnostic C shim provide interpreter callbacks, compiled
  built-in objectives, evaluation tracing, deterministic seeding of the
  library's lagged-Fibonacci generator, and pre-call validation that makes
  the packages memory-safe against every documented limit of the 1989
  working arrays. A Rust crate hosts both the translation and, via the
  same shim, the wrapped original; the Python extension exposes the two
  side by side as \code{backend="fortran"} and \code{backend="rust"}.
\item \textbf{A translation-fidelity study}
  (Section~\ref{sec:fidelity}). An early observation that ``wrapping
  Fortran is about $4\times$ faster than the Rust translation'' resisted
  the obvious explanation. By separating foreign-function call overhead,
  interpreter-callback cost, and algorithm-internal cost, we show the
  overheads are microseconds and nearly backend-independent, while the
  translated \code{BAYES1} planner was $1.7\times$ ($n{=}2$) to
  $12\times$ ($n{=}20$) slower because the port dropped two pruning
  rules of the original acquisition scan. Restoring them yields bitwise
  trajectory agreement with the 1989 binary and runtime parity. We also
  document independent correctness bugs found only because the original
  was executable: re-implemented Shekel and Hartmann test matrices were
  transposed relative to the Fortran \code{DATA} statements in both
  prior re-implementations (all five coefficient families in the pure-R
  port, two of five in the Rust port).
\item \textbf{A cross-ecosystem benchmark} (Sections~\ref{sec:bench}
  and~\ref{sec:results}). A COCO-style~\cite{hansen2021coco} suite (16
  test families, dimensions 2--10, 15 shifted instances each, budgets of
  $25n$ and $100n$ evaluations, identical bit-level instances in R and
  Python) comparing seven wrapped 1989 methods against SciPy
  (differential evolution~\cite{storn1997differential}, dual
  annealing~\cite{tsallis1996generalized,xiang1997generalized},
  DIRECT~\cite{jones1993lipschitzian,gablonsky2001locally},
  SHGO~\cite{endres2018shgo}), NLopt~\cite{johnson2007nlopt} (DIRECT-L,
  CRS2~\cite{kaelo2006some}, MLSL~\cite{rinnooykan1987stochastic},
  ISRES~\cite{runarsson2000stochastic}),
  CMA-ES~\cite{hansen2001completely}, Gaussian-process expected
  improvement (\pkg{scikit-optimize}~\cite{head2021skopt}), and the R
  packages
  GenSA~\cite{xiang2013gensa}, DEoptim~\cite{mullen2011deoptim}, and
  GA~\cite{scrucca2013ga}, evaluated with performance
  profiles~\cite{dolan2002benchmarking} and data
  profiles~\cite{more2009benchmarking}, and externally validated on the
  official BBOB suite of the COCO platform~\cite{hansen2021coco}
  (24 functions, 43{,}920 additional runs).
\item \textbf{A compound 1989 method} (\code{EXKOR/A2},
  Sections~\ref{sec:bench} and~\ref{sec:results}), answering the
  question whether the historic components compose into an improved
  method: LP-$\tau$ design, \code{ANAL2} eigenframe estimation, and
  \code{EXKOR} search, combined unchanged under a simple
  influence-ratio guard. The compound repairs four of five of
  \code{EXKOR}'s total-failure families and improves its mean rank in
  both ecosystems while conceding a statistical tie in aggregate solve
  rate---evidence that composition buys robustness rather than
  throughput, and that the 1989 interfaces' inability to warm-start
  taxes any composition.
\end{itemize}

Beyond the specific library, the paper argues a methodological point for
research software engineering: \emph{when modernizing numerical legacy
code, keep the original runnable and compare against it continuously}.
Every consequential defect we found---performance-relevant pruning rules
silently dropped, test-function coefficient matrices transposed, a
latent uninitialized-variable bug in the upstream double-precision
tree itself---was detected only through executable-to-executable
comparison, not code review. The lesson is timely rather than
antiquarian: large-language-model--assisted translation is currently
industrializing exactly the rewrite route studied here, at a scale
where nobody will hand-audit the output. Our failure taxonomy
(complexity-relevant control flow discarded as incidental; array
layouts transposed across language conventions; both invisible to
review and to unit tests of summary values) and the discipline that
catches it (full-trajectory parity against the executable original)
transfer directly to that setting.

\section{The GlobalMinimum library}
\label{sec:library}

\subsection{Methods}
\label{sec:methods}

We briefly describe the routines wrapped by our packages; the monograph
\cite{mockus1989bayesian} gives derivations, and
\cite{torn1989global,zilinskas2016stochastic} situate them historically.

\paragraph{One-step Bayesian search (\code{BAYES1}).}
After $L_0$ initial evaluations at LP-$\tau$ (Sobol-type) points, each
subsequent point maximizes a closed-form acquisition: for candidate $x$
with history $(x_i, y_i)_{i\le\ell}$ and incumbent $y^*$,
\begin{equation}
\alpha(x) \;=\; \min_{i\le \ell}\;
\frac{\lVert x - x_i\rVert^2}{\,y_i - y^* + \varepsilon\,},
\label{eq:fiap1}
\end{equation}
maximized over a Monte-Carlo candidate set of adaptive size
$m=\min(50n,\max(10n,\ell n))$ drawn from the library's additive
lagged-Fibonacci generator (\code{ATS}). Criterion~\eqref{eq:fiap1} is a
distance-to-merit ratio: a point is attractive when it is far from
previously evaluated points whose values are close to the incumbent. It
is the asymptotic simplification of the one-step Bayesian decision rule
derived in \cite{mockus1978application}, evaluable in $O(\ell n)$ without
any matrix algebra. Two pruning rules make the scan cheap in practice,
and both matter for Section~\ref{sec:fidelity}: the inner distance
accumulation aborts once the partial sum exceeds a threshold, and the
scan aborts entirely once $\alpha$ falls below the running second-best
candidate score (maintained across candidates in COMMON storage).

\paragraph{Wiener-model one-dimensional search (\code{EXTR},
\code{EXKOR}).} \code{EXTR} models a one-dimensional objective as a
Wiener process, evaluating where the conditional probability of improving
on the incumbent is largest, with a stopping rule at estimated posterior
confidence $0.95$; \code{EXKOR} applies it coordinate-wise. These
predate, and structurally resemble, modern one-dimensional
Bayesian-optimization schemes.

\paragraph{Multistart and deterministic methods.} \code{UNT} couples
uniform random starts with a Wiener-model uncertainty estimate to decide
which local minima to pursue; \code{GLOPT} clusters promising samples and
searches within clusters, in the spirit of
\cite{rinnooykan1987stochastic}; \code{LPMIN} searches LP-$\tau$
low-discrepancy designs, optionally after a factor-analysis phase that
orders variables by influence; \code{MIG1}/\code{MIG2} are Monte-Carlo
baselines. Local methods comprise a variable-metric quasi-Newton code
with finite-difference gradients (\code{MIVAR4}), a flexible-tolerance
constrained simplex (\code{FLEXI}), and recursive quadratic programming
(\code{REQP}). \code{ANAL1}/\code{ANAL2} screen variables by harmonic and
covariance-eigenstructure analysis of an evaluated design.

\subsection{Software archaeology}
\label{sec:archaeology}

The distribution contains a single-precision tree and a hand-converted
\code{REAL*8} tree. The conversion is instructive about pre-IEEE
portability practice: algorithm files received explicit
\code{IMPLICIT REAL*8} declarations, while the shared utilities
(\code{ATS}, \code{LPTAU}, machine constants, test objectives) remained
default-\code{REAL}, relying on an \code{f77 -r8}-style promotion flag.
On x87 hardware a \code{REAL} function result read as \code{REAL*8}
happened to work (results returned at extended precision on the
floating-point stack); on the x86-64 SSE ABI it silently returns garbage,
so a faithful modern build \emph{must} reproduce the promotion
(\code{-fdefault-real-8 -fdefault-double-8}) or convert the utilities
explicitly (we do both, converting the utility file and flagging the
rest). The audit also exposed genuine bugs in the upstream code itself.
\code{MIVAR4}'s double-precision line search renamed its
machine-constant bounds at the two sites where they are computed
(\code{SRMIN}${}\to{}$\code{SDMIN}) but not at the nine sites where they
are used, leaving the used variables implicitly zero; our vendored copy
restores the intended values. In \code{EXKOR}'s private copy of the
one-dimensional engine, a degenerate parabola fit on the first pass
through the local-optimization block stores the interpolation point
before it has ever been assigned---an uninitialized read flagged by
\code{-Wmaybe-uninitialized} during our determinism audit; tellingly,
the standalone \code{EXTR} contains the repaired form of the same
statement, so the defect was found and fixed upstream in one copy of
the duplicated code but not the other. Our vendored copies initialize
the point to the current simplex middle, which the degenerate branch
otherwise re-stores idempotently. Two pairs of files define identical
private helpers (\code{COR}/\code{ALNORM} in both \code{extr.f} and
\code{exkor.f}; \code{FAKTKK} in both \code{lpmin.f} and
\code{anal2.f}), which forbids linking the library into one shared
object without renaming---acceptable in 1989 where each program linked
one optimizer, fatal for a package that must load all of them.

Three interface properties shape any wrapper design: the objective
\code{FI(X,N)} and constraint routine \code{CONSTR(X,NX,R,K8)} are
\emph{link-time symbols}, not arguments; control parameters arrive
through offset-indexed \code{IPAR(30)}/\code{PAR(30)} arrays whose
validation failures print to a Fortran output unit and set a COMMON flag;
and all state, including the random-generator state and evaluation
histories capped at fixed sizes (dimension $\le 20$, $\le 1000$
evaluations for the Bayesian routines), lives in COMMON blocks.

\section{Package design}
\label{sec:packages}

\subsection{Architecture}

Both packages share a three-layer design (Figure~\ref{fig:arch}
schematically): the pristine vendored Fortran (four documented,
mechanically applied source patches: the \code{MIVAR4} repair, two
symbol renames, and one \code{DIMENSION} statement that listed scalars,
rejected by modern compilers); a Fortran trampoline defining \code{FI}
and \code{CONSTR} to forward to C entry points; and a language-agnostic
C shim that dispatches those calls to either a registered host-language
callback or one of the eight compiled built-in test objectives, while
counting evaluations, optionally recording the full evaluation trace,
and exposing get/set access to the 15-word \code{ATS} generator state.
Wrappers validate every documented limit \emph{before} entering Fortran,
so the upstream error branches---which would write to standard output
and, in 1989 style, poke a COMMON flag---are unreachable; the printing
parameter is pinned to silent.

Seeding deserves a note. The library initializes \code{ATS} from a
\code{BLOCK DATA} constant, so a fresh 1989 process was deterministic,
but state persisted \emph{across} calls within a process, making repeated
runs irreproducible. Our wrappers reset the state to the canonical
constants before each run by default---every call reproduces a fresh-process
run of the original code---and accept an integer seed that derives a
perturbed state through a Park--Miller generator, giving properly
replicated stochastic runs (the benchmark of Section~\ref{sec:bench}
uses this). We found and fixed the opposite defect in the repository's
earlier ad-hoc harness: optimizers silently ignored the seed, so
``30-seed'' summaries of the deterministic methods aggregated thirty
identical runs against genuinely varying competitors.

\begin{figure}[t]
\centering
\begin{lstlisting}
R / Python / Rust API      (validation, seeding, results, docs, tests)
        |  .Call / pyo3 / extern "C"
C shim  gm_shim.c          (callback registry, builtins, trace, evals, ATS)
        |  gmcall_ / gmcon_
Fortran trampoline gm_fi.f (FI, CONSTR -> C; GMATSE/GMATGE state access)
        |  link-time
GlobalMinimum real.8       (14 routines, vendored + 4 documented patches)
\end{lstlisting}
\caption{Layered wrapper architecture shared by the R package, the
Python extension, and the Rust crate.}
\label{fig:arch}
\end{figure}

\subsection{The R package}

\pkg{globalopt} for R compiles the Fortran into the package shared
object (standard \code{R CMD} toolchain; no external build steps). All
fourteen routines are exposed with roxygen documentation, a vignette,
and a testthat suite that pins LP-$\tau$ points, built-in objective
values, and full \code{BAYES1}/\code{MIG2} runs to reference values
obtained from a standalone C driver linked directly against the library.
Objectives are R closures (called through the shim with error trapping;
an R-level error or non-finite value aborts the search safely) or
built-in names (\code{"furasn"}, \code{"fush5"}, \ldots) that keep the
entire run in compiled code. Results carry the trace (points matrix,
values), the exact evaluation count from the shim, and the Fortran
termination code. A pure-R reference implementation of the core methods
is retained as \code{backend="reference"}---pedagogically useful, and
the measured base case for Section~\ref{sec:fidelity}'s interpreter
comparison. \code{R CMD check} passes with one expected note (the
compiled code retains Fortran \code{WRITE} statements on unreachable
error paths) and one portability warning for the promotion flags,
both documented.

\subsection{The Python package and Rust crate}

The Rust crate hosts the translation (\code{mig1}, \code{mig2},
\code{bayes1}, and modernized variants of the remaining methods) and,
behind a \code{fortran} feature, the wrapped original: \code{build.rs}
compiles the vendored sources with gfortran and links them with the same
C shim, exposing safe wrappers that validate limits, serialize calls
through a mutex (COMMON state is global), and return the same
result struct as the translation. The Python extension (pyo3/maturin)
publishes both sides as \code{backend="rust"} and
\code{backend="fortran"}; Python callables are invoked through identical
conversion machinery in either case, which is what makes the overhead
comparison of Section~\ref{sec:fidelity} clean. The wheel bundles
\code{libgfortran}, so end users need no Fortran toolchain.

\section{Wrap versus rewrite: a fidelity study}
\label{sec:fidelity}

\subsection{The ``4$\times$'' anomaly}

Ad-hoc timings in the project's history suggested the wrapped Fortran
\code{BAYES1} ran roughly four times faster than the Rust translation,
inviting explanations about foreign-function or interpreter overhead. We
decomposed the difference with three controlled experiments on the same
host (all scripts and raw data ship with the repository):

\begin{enumerate}
\item \textbf{Call and callback overhead} (algorithm cost
  $\approx 0$): \code{MIG2}, 1000 evaluations of the built-in
  \code{furasn} objective. Per evaluation, the wrapped Fortran costs
  0.5--1.7\,$\mu$s and the Rust translation 0.8--2.3\,$\mu$s across
  dimensions 2--20. Supplying the objective as a Python callable adds
  1--3\,$\mu$s per evaluation \emph{to either backend equally}; an R
  closure adds 2--4\,$\mu$s. No $4\times$ lives here.
\item \textbf{Algorithm-dominated runs}: \code{BAYES1}, 1000
  evaluations. Before our fix the Rust translation took 0.32\,s,
  3.4\,s, and 13.9\,s at $n=2,10,20$ against 0.18\,s, 1.02\,s, and
  1.16\,s for the wrapped Fortran measured in the same session---a
  \emph{dimension-dependent} $1.7\times$, $3.4\times$, $12\times$
  gap, incompatible with constant-per-call overhead.
\item \textbf{Interpreter base case}: the pure-R reference
  implementation of the same algorithm takes 55--200\,s, i.e.\
  $84\times$ ($n{=}10$, linearly extrapolated and therefore
  understated) to $465\times$ ($n{=}2$, measured directly) the wrapped
  Fortran, dwarfing every wrapping-related effect.
\end{enumerate}

\subsection{Diagnosis and repair}

The dimension dependence pointed into the acquisition
scan~\eqref{eq:fiap1}. The upstream \code{FIAP1} contains two pruning
rules that a line-by-line translation had discarded as incidental
control flow (\code{GOTO}-dense, COMMON-coupled): (i) the inner
squared-distance accumulation aborts as soon as the partial sum reaches
a threshold that certifies the observation cannot improve the score
--- the work saved grows linearly with dimension, matching the observed
scaling; and (ii) the whole history scan aborts when the score drops
below the running second-best across candidates, a bound communicated
via COMMON from the candidate loop (\code{MIG2F2}), which the
translation had replaced with a materialize-all-then-sort structure that
also allocated per candidate. Restoring both rules---by rewriting the
planner in the original's streaming two-best form---made the Rust
translation match the wrapped original \emph{bitwise} on 168 of 200
trace values on the reference problem (the remainder differ by one unit
in the last place, gfortran versus Rust standard-library cosine) with
runtime ratios of 1.31, 1.12, and 1.02 at $n=2,10,20$; the residual
low-dimension difference is per-candidate allocation, not algorithm.

\subsection{Correctness hazards of rewriting}

The same executable-to-executable audit caught quiet correctness bugs.
The Shekel and Hartmann coefficient matrices appear in the Fortran as
column-major \code{DATA} streams; both the pure-R and (for two of five
matrices) the Rust re-implementations had refilled them row-wise,
producing plausible-looking but wrong test functions---the transposed
Shekel-7 still has minima around $-10$ in the right region. Because the
compiled originals were callable next to the rewrites, a one-line test
(same trajectory under built-in versus re-implemented objective)
exposed the transpositions immediately. We take from this a simple
discipline for numerical modernization: \emph{keep the legacy binary
runnable inside the test suite; parity of full trajectories, not of
summary statistics, is the assertion of record.}

\section{Benchmark design}
\label{sec:bench}

\paragraph{Problems and instances.} Eight scalable families (sphere,
Rosenbrock, Rastrigin, Ackley, Griewank, L\'evy, Schwefel, Zakharov) at
$n\in\{2,5,10\}$ and eight fixed-dimension classics (Branin,
Goldstein--Price, six-hump camel, Shekel-5/7/10, Hartmann-3/6)
\cite{surjanovic2013library}, each in 15 instances formed by seeded
coordinate shifts of up to $\pm5\%$ of the box (COCO
style~\cite{hansen2021coco}), normalized so every instance has optimal
value exactly~0. Instance generation uses a Park--Miller stream shared
bit-for-bit by the R and Python harnesses, verified identical across
languages.

\paragraph{Protocol.} Budgets of $25n$ and $100n$ evaluations. Every
solver sees the objective only through a recording wrapper that counts
evaluations, tracks the best value within budget, and records the first
evaluation index reaching gap tolerances $10^{-1},\dots,10^{-8}$;
solvers that overrun their nominal budget are truncated at analysis time
(SciPy's SHGO is the only material offender). One run per (method,
instance, budget); variation across the 15 instances provides
replication, with stochastic solvers additionally seeded per instance
(including the wrapped 1989 methods, through the ATS state described in
Section~\ref{sec:packages}).

\paragraph{Methods.} Python: seven wrapped 1989 methods
(\code{BAYES1}, \code{MIG2}, \code{GLOPT}, \code{UNT}, \code{LPMIN},
\code{EXKOR}, \code{LBAYES}), the repaired Rust translations of
\code{BAYES1}/\code{MIG2}, SciPy differential evolution, dual annealing,
DIRECT, SHGO, NLopt DIRECT-L, CRS2, MLSL, ISRES, CMA-ES
(\pkg{pycma}), Gaussian-process Bayesian optimization with expected
improvement (\pkg{scikit-optimize}; the modern realization of the
program \code{BAYES1} approximates), and uniform random search as
control. The GP baseline runs only on the $25n$ tier at $n\le 6$: its
cubic-cost model refits already average tens of seconds per run there,
which is also its natural operating regime. R: the same seven
wrapped methods, GenSA, DEoptim, GA, NLopt via \pkg{nloptr} (DIRECT-L,
CRS2, ISRES), and random search. The wrapped methods inherit the 1989
array caps (e.g.\ 1000 evaluations for \code{BAYES1}, 500 for
\code{UNT}/\code{EXKOR}); where the budget exceeds a cap the method
simply stops early and its profile flattens, which we report as-is:
the caps are part of the artifact under study.

\paragraph{Metrics.} Dolan--Mor\'e performance
profiles~\cite{dolan2002benchmarking} on evaluations-to-target and
Mor\'e--Wild data profiles~\cite{more2009benchmarking} at gap tolerances
$10^{-2}$ and $10^{-4}$; median terminal gap tables; wall-clock time per
run (cheap objectives, so time exposes solver-internal and FFI cost);
and the FFI microbenchmarks of Section~\ref{sec:fidelity}. Profiles and
rank summaries are descriptive; for the six headline pairwise
comparisons quoted in Section~\ref{sec:results} we additionally run
exact McNemar tests on the paired solved/unsolved outcomes over the 480
instances (all quoted $p$-values remain significant, or not, under
Holm--Bonferroni correction across the six planned tests).

\paragraph{A compound 1989 method (EXKOR/A2).} The component
measurements pose a research question: \emph{can the historic
components be composed, unchanged, into an improved method?} The
library supplies the parts: LP-$\tau$ designs, \code{ANAL2}'s
eigenframe estimate, and \code{EXKOR}'s coordinate search.
\code{EXKOR/A2} spends 10\% of the budget on an LP-$\tau$ design,
estimates the eigenframe with \code{ANAL2}, runs plain \code{EXKOR}
on the full box for 45\%, and spends the final 45\% restarting
\code{EXKOR} at the incumbent---in the rotated frame when the dominant
eigen-direction explains at least $1.3\times$ the influence of the best
original axis, in the original axes otherwise. The guard matters: an
initial always-rotate variant repaired \code{EXKOR}'s failures on
linearly coupled problems but destroyed separable ones (rotating a
separable function makes it non-separable), and a fixed 25\% design
fraction taxed every problem; we report the portfolio form above and
disclose that it is the second of two design iterations, both scripted
in the repository.

\paragraph{BBOB validation.} Because a fixed collection of classical
test functions can encode unintended structure (four of our families
even ship with the library), we replicate the Python-side experiment on
the official BBOB suite via \pkg{cocoex}~\cite{hansen2021coco}: all 24
functions, dimensions $\{2,5,10\}$, instances 1--15 (1{,}080 problem
instances), the same recorder protocol, budget tiers, seeding, and
method set. Instance optima are recovered from the suite's optimizer
locations so the same gap-based analysis applies unchanged; runs were
sharded across three hosts.

\section{Results}
\label{sec:results}

The sweep comprises 19{,}560 Python runs (including the 360-run
GP-Bayesian-optimization tier) and 14{,}400 R runs on the classical
suite, plus 43{,}920 BBOB runs (Section~\ref{sec:bbob}); no run failed
except the intentional SHGO exclusions above $n=5$ --- where its
simplicial machinery takes minutes and gigabytes per run --- which are
recorded as failures and count against it.
Raw CSVs, the analysis script, and the exact numbers below ship with the
repository.

\begin{figure}[t]
\centering
\includegraphics[width=0.49\textwidth]{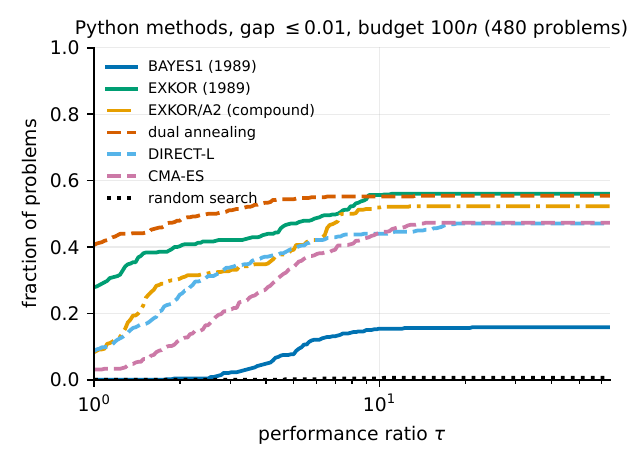}%
\includegraphics[width=0.49\textwidth]{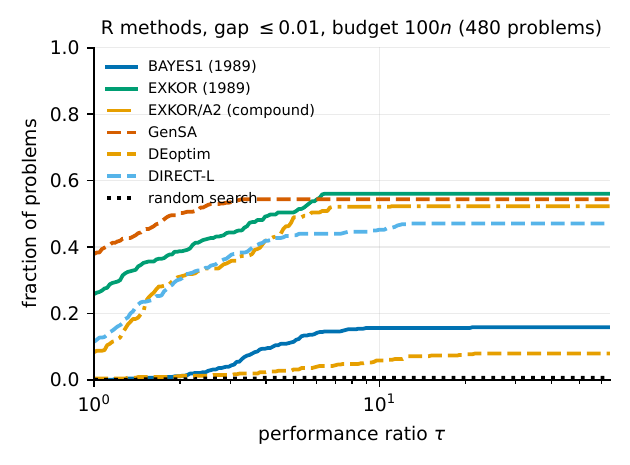}
\caption{Dolan--Mor\'e performance profiles on evaluations to reach gap
$10^{-2}$ within the $100n$ budget (480 problem instances; unsolved
instances count in the denominator). Solid lines: wrapped 1989 methods;
dashed: modern competitors; dotted: random search.}
\label{fig:perf}
\end{figure}

\begin{figure}[t]
\centering
\includegraphics[width=0.49\textwidth]{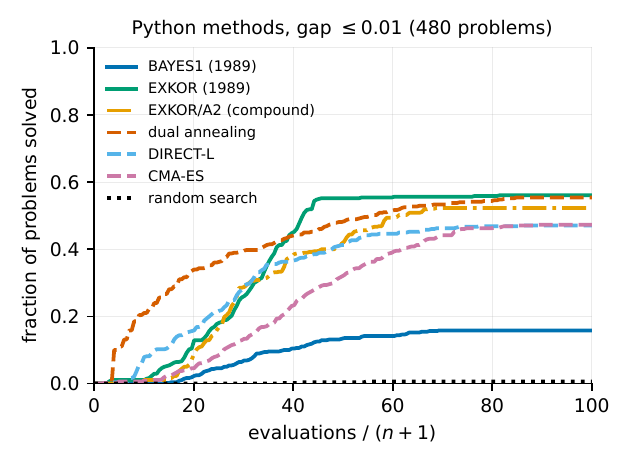}%
\includegraphics[width=0.49\textwidth]{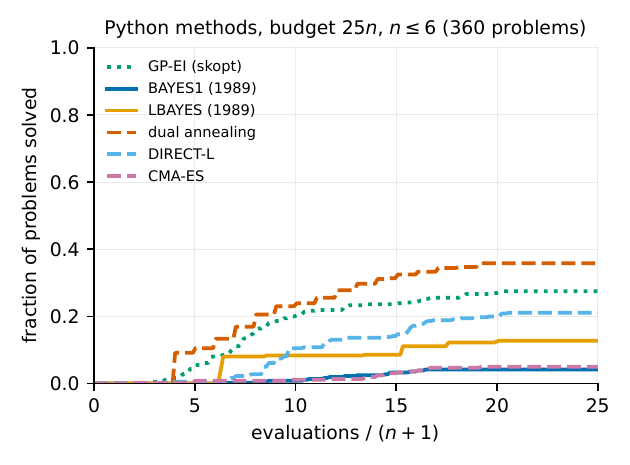}
\caption{Mor\'e--Wild data profiles (fraction of instances solved to
gap $10^{-2}$ versus evaluations$/(n{+}1)$), Python methods. Left: the
$100n$ budget. Right: the $25n$ budget restricted to $n\le 6$, the
tier on which the GP-EI baseline runs.}
\label{fig:data}
\end{figure}

\begin{figure}[t]
\centering
\includegraphics[width=0.6\textwidth]{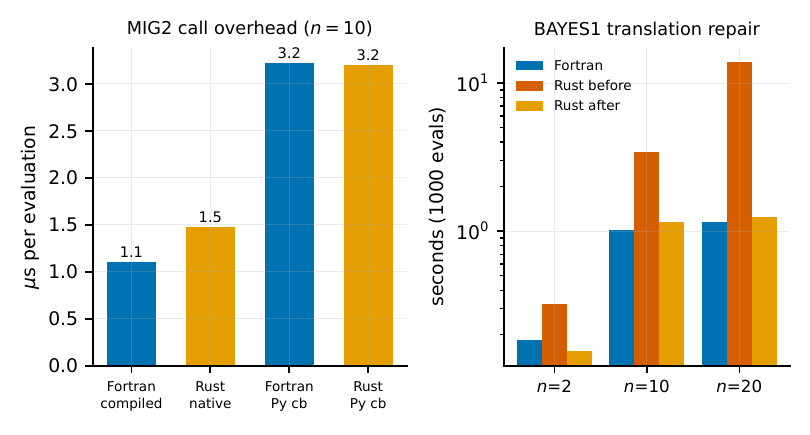}
\caption{The overhead/fidelity decomposition of
Section~\ref{sec:fidelity}: per-evaluation cost of the
overhead-dominated \code{MIG2} ($n{=}10$), and \code{BAYES1} runtime
before/after the pruning repair.}
\label{fig:ffi}
\end{figure}

\subsection{Where the 1989 methods stand}

Figure~\ref{fig:perf} shows performance profiles at gap $10^{-2}$ under
the $100n$ budget; Figure~\ref{fig:data} (left) the corresponding data
profile. Three regimes are visible.

\paragraph{Generous budgets ($100n$).} The wrapped methods produce
bitwise-identical sweeps in R and Python (same solve sets, discordant
counts, and $p$-values --- itself a strong wrapper-fidelity check), so
we quote their numbers once. The coordinate Wiener-model search
\code{EXKOR} is the strongest pure 1989 method and competitive with
everything we ran: it solves 56.0\% of instances within budget ---
statistically tied with SHGO (56.2\%) and dual annealing (55.4\%; 140
versus 137 discordant instances, McNemar $p=0.90$), tied with GenSA
(54.4\%, 133 versus 125, $p=0.66$), and ahead of CMA-ES (47.3\%,
$p=0.006$) and DIRECT-L (47.1\%, $p=0.003$, identically in both
ecosystems). Tightening the tolerance to $10^{-4}$ preserves the
picture (SHGO 54.2\%, dual annealing 52.9\%, \code{EXKOR} 50.2\%).
\code{EXKOR}'s profile is sharply bimodal across problem families,
exactly as its structure predicts: on separable or coordinate-aligned
landscapes it is essentially perfect (100\% of Hartmann-6, L\'evy,
Schwefel, six-hump camel, and sphere instances; 91\% Rastrigin; 82\%
Ackley), while on coupled valleys it solves nothing (0\% on
Rosenbrock, Branin, Goldstein--Price, and Zakharov), and on Hartmann-3
its coordinate-wise moves descend into the secondary basin on every
instance (a constant terminal gap of 0.773). Averaged ranks over all
problem groups therefore place it eighth of twenty in Python (mean
rank 6.7 versus 4.8 for DIRECT-L) and fourth of fifteen in R (4.8,
behind GenSA at 2.7, DIRECT-L at 3.4, and its own compound variant),
ahead of CRS2, ISRES, DEoptim, and GA in both ecosystems.

\paragraph{Tight budgets ($25n$).} Small-budget behavior reverses the
order: dual annealing (31.0\%) and GenSA (28.5\%) dominate, the
multistart \code{LBAYES} is the best 1989 method (12.7\%, third
overall in R at this budget), and \code{EXKOR} (4.4\%) has not yet
completed its first coordinate cycles. The one-step Bayesian
\code{BAYES1} sits mid-field at both budgets (15.8\% at $100n$); its
fixed candidate-sampling acquisition is dominated by modern annealing
schedules on these landscapes, though it beats CRS2, ISRES, and the
differential-evolution and genetic-algorithm packages ---
implementations published one to two decades later --- under the same
protocol.

\paragraph{Against modern Bayesian optimization.} On the tier where
GP-based expected improvement is computationally feasible ($25n$
budget, $n\le 6$; 360 instances), it solves 27.5\% of instances ---
decisively ahead of its 1989 ancestors \code{BAYES1} (4.2\%, McNemar
$p<10^{-21}$) and \code{LBAYES} (12.8\%, $p<10^{-10}$), and behind
only dual annealing (35.8\%, $p=8\times10^{-4}$). Figure~\ref{fig:data}
(right) shows the full data profile. The quality gap is what
thirty-five years of surrogate and acquisition development bought; the
price is solver-internal cost: median 77\,s per $25n$-budget run for
GP-EI versus 9\,ms for \code{BAYES1} on the same tier --- nearly four
orders of magnitude --- which is exactly the
cheap-acquisition/many-evaluations trade the 1989 design targeted.

\paragraph{The compound method.} \code{EXKOR/A2} answers its research
question with a precise ``yes, for robustness --- no, for aggregate
throughput.'' At $100n$ it repairs four of plain \code{EXKOR}'s five
zero-solve families (Hartmann-3 $0\to100\%$, Zakharov $0\to33\%$,
Goldstein--Price $0\to27\%$, Rosenbrock $0\to13\%$; only Branin
resists), cutting zero-solve families from five to two, and improves
the all-problems mean rank in both ecosystems (6.3 versus 6.7 in
Python; 4.2 versus 4.8 in R, where it overtakes plain \code{EXKOR} for
third place). The price is paid exactly where the budget split
predicts: on multimodal separable problems that plain \code{EXKOR}
solved with its full budget (Rastrigin $-78$, six-hump camel $-60$,
Schwefel $-24$ percentage points), for an aggregate solve rate of
52.3\% versus 56.0\% (41 versus 59 discordant, $p=0.089$ --- a
statistical tie, identical in both languages). On BBOB it is nominally
ahead (15.3\% versus 14.6\%, $p=0.32$). At the $25n$ tier its
three-phase structure starves every phase and it should not be used.
The composition also surfaces a structural tax: the 1989 interfaces
cannot warm-start from prior evaluations, so each phase re-buys
information the process already paid for.

\paragraph{Deployed fidelity check.} The repaired Rust translation of
\code{BAYES1} is statistically indistinguishable from the wrapped
original across the sweep: solve rates 15.6\% versus 15.8\% and only
7 of 480 instances with discordant outcomes (McNemar $p=1$; 23 of
1{,}080 on BBOB, $p=0.21$)---the benchmark-scale confirmation of the
trace-level parity of Section~\ref{sec:fidelity}. The wrapped Fortran
methods themselves replicate bitwise between the R and Python harnesses
on identical instances, which is what exposed both the upstream
uninitialized-variable defect of Section~\ref{sec:archaeology} and the
harness defect discussed under threats.

\paragraph{Baselines and sanity checks.} Random search solves 0.6\% of
instances at $100n$; every method clears it decisively. Differential
evolution and GA suffer from population sizing at these budgets
($10n$--$50$ individuals leave few generations); we report them under
the same budget-scaled defaults as everything else rather than tuning
per problem.

\subsection{External validation on BBOB}
\label{sec:bbob}

\begin{figure}[t]
\centering
\includegraphics[width=0.49\textwidth]{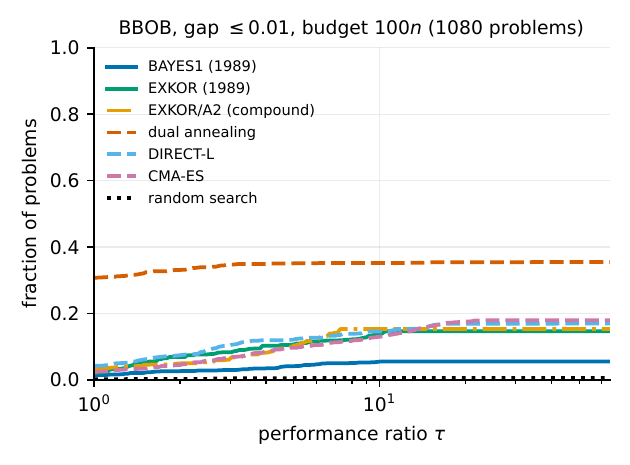}%
\includegraphics[width=0.49\textwidth]{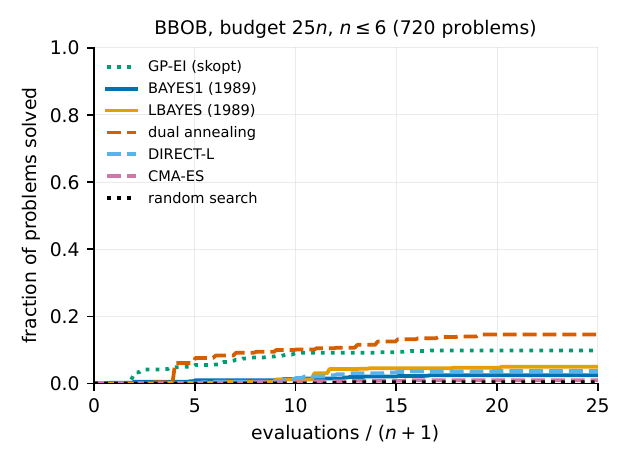}
\caption{BBOB suite (24 functions, $n\in\{2,5,10\}$, 15 instances).
Left: performance profiles at gap $10^{-2}$, budget $100n$. Right:
tight-budget data profile ($25n$, $n\le6$) including the GP-EI
baseline.}
\label{fig:bbob}
\end{figure}

The BBOB replication (Figure~\ref{fig:bbob}) separates what
generalizes from what was structure-specific. Three findings
replicate. Dual annealing remains the overall winner (35.5\% of
instances at $100n$, ahead of SHGO at 30.0\%; 14.6\% at $25n$,
$n\le6$). GP-EI remains decisively ahead of its 1989 ancestors at
tight budgets (9.9\% versus 5.3\% for \code{EXKOR/A2} and 5.0\% for
\code{LBAYES}). And the repaired Rust translation of \code{BAYES1}
remains statistically indistinguishable from the wrapped Fortran (23
of 1{,}080 instances discordant, $p=0.21$). One finding is properly
cut down to size: \code{EXKOR}, statistically tied with dual annealing
on the classical set, drops to sixth (14.6\%, versus 252--27
discordant instances in dual annealing's favor, $p<10^{-15}$) ---
BBOB's deliberately non-separable, ill-conditioned function groups
remove exactly the coordinate-aligned structure the classical test
canon is rich in. The compound \code{EXKOR/A2} edges past its parent
here (15.3\%, 22 versus 15 discordant, $p=0.32$), consistent with its
design target; \code{EXKOR} remains slightly behind DIRECT-L (65
versus 91, $p=0.045$) and ahead of CRS2, ISRES, and differential
evolution; \code{BAYES1} solves 5.6\%. We read the pair of suites as
bracketing practice: real engineering objectives fall between the
classical canon's separability and BBOB's adversarial conditioning,
and the 1989 methods' standing moves accordingly.

\subsection{Overheads and throughput}

With cheap analytic objectives, wall time exposes solver internals.
The wrapped 1989 methods occupy the fastest tier: median time per
$100n$-budget run is 1.8\,ms for \code{EXKOR}, 3.8\,ms for the
compound \code{EXKOR/A2}, 2.5--2.7\,ms for \code{MIG2} and
\code{GLOPT}---at or below the 2.8\,ms of a bare random-search loop
in Python, i.e.\ below the interpreter cost of the objective itself.
NLopt methods run at 5--7\,ms, SciPy DIRECT at 7\,ms, while dual
annealing (38\,ms), differential evolution (42\,ms), and CMA-ES
(107\,ms) pay their library-internal overheads, and \code{BAYES1}
(79\,ms Python, 102\,ms R) pays its $O(\ell^2 n)$ acquisition scan.
Combined with Section~\ref{sec:fidelity}: for objectives costing under
a microsecond the callback into the host language dominates everything
else, and the built-in compiled-objective path of our packages removes
even that.

\section{Discussion}
\label{sec:discussion}

\paragraph{Wrap or rewrite?} Our evidence favors \emph{wrap first,
rewrite selectively, and never rewrite without the wrapped original in
the loop}. Wrapping delivered the genuine algorithms at compiled speed
in weeks of effort dominated by interface archaeology, not numerics;
the fidelity incidents of Section~\ref{sec:fidelity} were each invisible
to code review and immediate under trajectory-parity testing. The
repaired translation is not redundant: it removes the 1989 array caps
and COMMON-block thread-unsafety, and compiles to platforms without a
Fortran toolchain. The two routes are complements---but only the
executable original can arbitrate the translation's fidelity.

\paragraph{Legacy algorithms, modern context.} The acquisition
criterion~\eqref{eq:fiap1} costs $O(\ell n)$ per candidate against the
$O(\ell^3)$ of GP-based acquisition, and the measured throughput
reflects it: the wrapped methods run at or below the cost of a bare
interpreter loop over the objective. The benchmark places the 1989
methods squarely in the modern mid-field---\code{EXKOR} at the top on
separable landscapes and even overall at the $100n$ budget,
\code{LBAYES} respectable at tight budgets, \code{BAYES1} ahead of
several methods a generation younger---while GenSA and dual annealing,
which postdate the library by a decade or more, are the consistent
overall winners. For practitioners the packages are thus more than a
museum: on separable or near-separable problems \code{EXKOR} is a
legitimately strong, extremely cheap choice, and the library's
LP-$\tau$ designs and screening routines remain useful building
blocks. Against its direct descendant, GP-based expected improvement,
the comparison quantifies the field's progress cleanly: the modern
acquisition solves $6.5\times$ as many tight-budget instances as
\code{BAYES1} while spending nearly four orders of magnitude more time
per run---each side of that trade still has its regime.

\paragraph{Scope: what the 1989 artifact cannot address.} Our
experiments stop at $n=10$ continuous box-constrained problems because
the artifact does: the working arrays cap most methods at $n\le 20$
and around $10^3$ evaluations, and the interfaces know nothing of
categorical, conditional, or batch-parallel evaluation. The dominant
global-optimization workload of the past decade---configuration and
hyperparameter search for deep networks---is high-dimensional, mixed
discrete-continuous, conditional, and noisy, and is served by purpose-built
machinery such as TPE and SMAC~\cite{bergstra2011algorithms,hutter2011smac};
wrapping a 1989 library cannot and should not compete there, and we
make no such claim. Two historical footnotes are worth recording,
though: \code{LBAYES} already supports leading integer variables
(handled by interpolation between adjacent integer points), and its
stochastic-approximation step schedule was designed for noisy
objectives---the concerns of modern configuration search were visible
in 1989, at 1989 scale.

\paragraph{Recommendations.} The measured evidence supports concrete
guidance. (1)~Under generous budgets on cheap continuous problems,
dual annealing (Python) and GenSA (R) are the strongest defaults we
measured; DIRECT-L is the strongest deterministic choice. (2)~Under
tight budgets with expensive objectives, GP-based expected improvement
is worth its solver cost as long as one solver step is cheaper than
one objective evaluation; below that, dual annealing again.
(3)~On problems known or suspected to be separable or coordinate-aligned,
\code{EXKOR} delivers top-tier quality at the lowest solver cost we
measured (below a bare interpreter loop)---and its failure mode is
legible: it loses exactly when the landscape couples coordinates. If
total failures are costlier than average regressions, its compound
variant \code{EXKOR/A2} trades a few points of aggregate solve rate
for substantially fewer zero-solve families.
(4)~For modernizers: wrap first; translate only with the original in
the test loop asserting full-trajectory parity; treat every branch of
legacy control flow as load-bearing until proven otherwise.

\paragraph{Toward compound algorithms.} The compound question is
answered quantitatively in Section~\ref{sec:results}: composing the
library's own parts (LP-$\tau$ + \code{ANAL2} + \code{EXKOR} under
an influence-ratio guard) buys robustness --- four of five total-failure
families repaired, better mean ranks in both ecosystems, a nominal edge
on BBOB --- but not aggregate throughput, and the design iterations
that led to the guard and the portfolio split are themselves evidence
of how easily naive composition destroys the very structure a
coordinate method exploits. A second composition we have not yet built
follows the same logic at the acquisition level: use the $O(\ell n)$
criterion~\eqref{eq:fiap1} to prescreen thousands of candidates and
reserve the $O(\ell^3)$ Gaussian-process posterior for the
shortlist --- a budget-matched bridge between the 1989 trade and the
modern one. The deeper obstacle either way is interface, not
mathematics: the 1989 routines cannot ingest prior evaluations, so
every composed phase re-purchases information the process already
holds.

\paragraph{Threats to validity.} Single host, cheap analytic
objectives (time comparisons expose overheads, not expensive-objective
regimes); default-leaning hyperparameters for competitors (budget-scaled
population sizes; no per-problem tuning for any method, including
ours---the wrapped methods use the upstream example parameters with
budget-scaled evaluation counts); the 1989 caps censor high-budget
behavior of the wrapped methods; microsecond-scale timings taken with
background load and reported as medians of repeated runs (an
idle-machine replication on a second host reproduced the post-repair
Rust/Fortran ratios at 1.26/1.16/1.10 for $n=2/10/20$). Four of the
sixteen classical test families (the Shekel and Hartmann variants)
also ship as demonstration objectives with the 1989 library, a
potential selection bias in its favor; the seeded shifts perturb them
away from the shipped coefficients, and the BBOB replication
(Section~\ref{sec:results}) quantifies exactly how the conclusions
move on an independently designed suite. One run per instance follows
COCO practice; the paired tests operate on the 480 (custom) and
1{,}080 (BBOB) instances rather than on within-instance replicates.
Finally, our own harness supplied a cautionary tale of exactly the kind
this paper documents: an early version of the R driver used \code{pi}
as a loop index, silently shadowing R's built-in $\pi$ constant and
corrupting every trigonometric test objective evaluated after the first
iteration---the sweep remained internally consistent (all methods saw
the same corrupted functions) and was exposed only by the
cross-language identity checks, after which all R results were
regenerated. Redundant implementations that must agree bitwise catch
what internal consistency cannot.

\section{Related work}
\label{sec:related}

Global-optimization benchmarking methodology is standardized by
performance profiles~\cite{dolan2002benchmarking}, data
profiles~\cite{more2009benchmarking}, and the COCO
platform~\cite{hansen2021coco}; we follow all three. The historical
position of the Vilnius school is surveyed
in~\cite{torn1989global,zilinskas2016stochastic}; modern Bayesian
optimization in~\cite{shahriari2016taking,frazier2018tutorial}. The
software route of wrapping venerable Fortran into scripting ecosystems
is the founding pattern of scientific Python and
R~\cite{virtanen2020scipy,chambers2008software}; our contribution is a
measured head-to-head of that pattern against a memory-safe-language
rewrite of the same code, with the failure modes documented.

\section{Availability}
\label{sec:availability}

The R package, Python wheel sources, Rust crate, vendored and patched
Fortran, benchmark harnesses, raw results, and every script producing a
number or figure in this paper are available at
\url{https://github.com/audrism/globalopt}. The upstream distribution
remains available from~\cite{globalminimumfortran} and is vendored
verbatim (patches applied mechanically and documented) for
reproducibility.

\paragraph{Acknowledgments.} The original MINIMUM/GlobalMinimum library
is the work of Jonas Mockus (1931--2021), the author's father; this
paper is dedicated to his memory. The familial relationship is
disclosed in the interest of transparency; all comparative results are
produced by the scripted, seeded pipeline shipped with the repository.

\appendix
\section{Median terminal gaps}
\label{app:gaps}

Table~\ref{tab:gaps} lists median terminal gaps at the $100n$ budget by
method: over the scalable families at each dimension, and over the
fixed-dimension classics. ``--'' marks methods not run in that setting
(SHGO above $n=5$).

\begin{table}[h]
\centering\small
\begin{tabular}{llrrrr}
\toprule
Ecosystem & Method & \multicolumn{3}{c}{scalable, median gap at $100n$} & classics \\
 & & $n{=}2$ & $n{=}5$ & $n{=}10$ & \\
\midrule
Python & dual annealing & 2.1e-12 & 1.7e-1 & 1.1e-1 & 6.5e-13 \\
Python & DIRECT-L & 6.1e-6 & 9.8e-1 & 4.1e+00 & 2.4e-6 \\
Python & CMA-ES & 4.0e-5 & 1.1e-1 & 5.1e-1 & 4.5e-5 \\
Python & EXKOR (1989) & 4.6e-5 & 2.5e-5 & 1.9e-5 & 7.7e-1 \\
Python & SHGO & 4.6e-5 & 1.6e+00 & -- & 3.2e-14 \\
Python & DIRECT (SciPy) & 1.3e-4 & 1.2e+00 & 6.0e+00 & 4.8e-5 \\
Python & MLSL & 2.4e-3 & 1.3e+00 & 2.0e+00 & 2.7e-4 \\
Python & EXKOR/A2 (compound) & 6.0e-3 & 1.9e-1 & 1.2e-1 & 6.6e-3 \\
Python & CRS2 & 5.3e-2 & 3.3e+00 & 8.0e+00 & 2.1e-1 \\
Python & diff. evolution & 1.1e-1 & 7.2e+00 & 3.7e+01 & 4.2e-1 \\
Python & GLOPT (1989) & 1.3e-1 & 3.4e+00 & 9.5e+00 & 2.6e-1 \\
Python & BAYES1 (1989) & 1.5e-1 & 6.9e+00 & 1.2e+01 & 3.5e-1 \\
Python & BAYES1 (Rust) & 1.5e-1 & 7.3e+00 & 1.2e+01 & 3.9e-1 \\
Python & MIG2 (Rust) & 2.6e-1 & 1.8e+01 & 6.8e+01 & 8.2e-1 \\
Python & LPMIN (1989) & 3.2e-1 & 7.1e+00 & 1.2e+01 & 1.0e+00 \\
Python & LBAYES (1989) & 3.4e-1 & 1.1e+00 & 2.8e+00 & 3.2e-1 \\
Python & UNT (1989) & 3.6e-1 & 1.5e+01 & 7.7e+01 & 1.1e+00 \\
Python & ISRES & 4.3e-1 & 6.9e+00 & 1.2e+01 & 9.2e-1 \\
Python & MIG2 (1989) & 5.6e-1 & 1.7e+01 & 7.4e+01 & 1.0e+00 \\
Python & random search & 6.2e-1 & 1.6e+01 & 7.1e+01 & 1.5e+00 \\
\midrule
R & GenSA & 5.3e-15 & 9.0e-2 & 2.7e-1 & 3.8e-15 \\
R & DIRECT-L & 6.6e-6 & 9.8e-1 & 4.1e+00 & 2.7e-6 \\
R & EXKOR (1989) & 4.6e-5 & 2.4e-5 & 1.9e-5 & 7.7e-1 \\
R & EXKOR/A2 (compound) & 6.0e-3 & 2.5e-1 & 1.2e-1 & 6.6e-3 \\
R & CRS2 & 4.8e-2 & 3.5e+00 & 9.4e+00 & 2.1e-1 \\
R & DEoptim & 1.1e-1 & 9.5e+00 & 4.5e+01 & 5.8e-1 \\
R & GLOPT (1989) & 1.3e-1 & 3.4e+00 & 9.5e+00 & 2.6e-1 \\
R & BAYES1 (1989) & 1.5e-1 & 6.9e+00 & 1.2e+01 & 3.5e-1 \\
R & LPMIN (1989) & 3.2e-1 & 7.1e+00 & 1.2e+01 & 1.0e+00 \\
R & GA & 3.2e-1 & 4.8e+00 & 8.5e+00 & 7.8e-1 \\
R & LBAYES (1989) & 3.4e-1 & 1.1e+00 & 2.8e+00 & 3.2e-1 \\
R & UNT (1989) & 3.6e-1 & 1.5e+01 & 7.7e+01 & 1.1e+00 \\
R & ISRES & 4.0e-1 & 7.2e+00 & 1.2e+01 & 9.1e-1 \\
R & MIG2 (1989) & 5.6e-1 & 1.7e+01 & 7.4e+01 & 1.0e+00 \\
R & random search & 7.0e-1 & 1.6e+01 & 7.9e+01 & 1.0e+00 \\
\bottomrule
\end{tabular}
\caption{Median gap to the global optimum at budget $100n$.}
\label{tab:gaps}
\end{table}

\bibliographystyle{unsrtnat}
\bibliography{refs}

\end{document}